%% jan. 21, 2002
%% jan. 23, 2002
%% july 1, 2002
%% minor update, august 20, 2002. 

%%% converted to small note for Dyson CMP volume. June 9, 2004
%%%
%%% important updates july 15, 2004

\input harvmac.tex

\lref\szenes{A. Szenes, ``The combinatorics of the Verlinde
formulas,''  alg-geom/9402003.}

\lref\rt{Russo-Tseytlin 
hep-th/9508068.}
 
\lref\takayanagi{T. Takayanagi and T. Uesugi, ``D-branes in Melvin 
background,'' hep-th/0110200.}

\lref\mourad{C. Angelantonj, E. Dudas, and J. Mourad, ``Orientifolds 
of String Theory Melvin backgrounds,'' hep-th/0205096.}

%\LiuKB
\lref\LiuKB{
H.~Liu, G.~Moore and N.~Seiberg,
``Strings in time-dependent orbifolds,''
JHEP {\bf 0210}, 031 (2002)
[arXiv:hep-th/0206182].
%%CITATION = HEP-TH 0206182;%%
}

%\AdamsSV
\lref\aps{
A.~Adams, J.~Polchinski and E.~Silverstein,
``Don't panic! Closed string tachyons in ALE space-times,''
JHEP {\bf 0110}, 029 (2001)
[arXiv:hep-th/0108075].
%%CITATION = HEP-TH 0108075;%%
}

%\HarveyWM
\lref\HarveyWM{
J.~A.~Harvey, D.~Kutasov, E.~J.~Martinec and G.~Moore,
``Localized tachyons and RG flows,''
arXiv:hep-th/0111154.
%%CITATION = HEP-TH 0111154;%%
}

%\DixonQV
\lref\DixonQV{
L.~J.~Dixon, D.~Friedan, E.~J.~Martinec and S.~H.~Shenker,
``The Conformal Field Theory Of Orbifolds,''
Nucl.\ Phys.\ B {\bf 282}, 13 (1987).
%%CITATION = NUPHA,B282,13;%%
}

%\SuyamaGD 
\lref\SuyamaGD{ 
T.~Suyama, 
``Properties of string theory on Kaluza-Klein Melvin background,'' 
hep-th/0110077. 
%%CITATION = HEP-TH 0110077;%% 
} 
 
%\RussoNA 
\lref\RussoNA{ 
J.~G.~Russo and A.~A.~Tseytlin, 
``Supersymmetric fluxbrane intersections and closed string tachyons,'' 
hep-th/0110107. 
%%CITATION = HEP-TH 0110107;%% 
}

\lref\marcolli{M. Marcolli, ``Modular curves, C* algebras, and 
chaotic cosmology,'' arXiv:math-ph/0312035;
M. Marcolli, ``Limiting modular symbols and the Lyapunov spectrum,'' 
arXiv:math.NT/0111093; Y. Manin and M. Marcolli, 
``Continued fractions, modular symbols, and non-commutative geometry,'' 
arXiv:math.NT/0102006.}

 %\HeadrickHZ
\lref\HeadrickHZ{
M.~Headrick, S.~Minwalla and T.~Takayanagi,
``Closed string tachyon condensation: An overview,''
arXiv:hep-th/0405064.
%%CITATION = HEP-TH 0405064;%%
}

%\MooreFG
\lref\MooreFG{
G.~W.~Moore,
``Les Houches lectures on strings and arithmetic,''
arXiv:hep-th/0401049.
%%CITATION = HEP-TH 0401049;%%
}

%\DavidVM
\lref\DavidVM{
J.~R.~David, M.~Gutperle, M.~Headrick and S.~Minwalla,
``Closed string tachyon condensation on twisted circles,''
JHEP {\bf 0202}, 041 (2002)
[arXiv:hep-th/0111212].
%%CITATION = HEP-TH 0111212;%%
}

%\MartinecTZ
\lref\MartinecTZ{
E.~J.~Martinec,
``Defects, decay, and dissipated states,''
arXiv:hep-th/0210231.
%%CITATION = HEP-TH 0210231;%%
}

%\DijkgraafTF
\lref\DijkgraafTF{
R.~Dijkgraaf and E.~Verlinde,
``Modular Invariance And The Fusion Algebra,''
Nucl.\ Phys.\ Proc.\ Suppl.\  {\bf 5B}, 87 (1988).
%%CITATION = NUPHZ,5B,87;%%
}
%\HarveyGQ
\lref\HarveyGQ{
J.~A.~Harvey, S.~Kachru, G.~W.~Moore and E.~Silverstein,
``Tension is dimension,''
JHEP {\bf 0003}, 001 (2000)
[arXiv:hep-th/9909072].
%%CITATION = HEP-TH 9909072;%%
}

\lref\schmidt{W.M. Schmidt, {\it Diophantine Approximation}, Springer-Verlag, LNM 785;
{\it Diophantine approximations and diophantine equations}, Springer-Verlag LNM 1467.}

\lref\pollicott{M. Pollicott and H. Weiss, ``Multifractal analysis of Lyapunov exponent for 
continued fraction and Manneville-Pomeau transformations and applications to 
Diophantine approximation,'' Commun. Math. Phys. {\bf 207}(1999)145.}

%\GutperleMB
\lref\GutperleMB{
M.~Gutperle and A.~Strominger,
``Fluxbranes in string theory,''
JHEP {\bf 0106}, 035 (2001)
[arXiv:hep-th/0104136].
%%CITATION = HEP-TH 0104136;%%
}

\lref\cassels{J. Cassels, {\it An Introduction to Diophantine Approximation}, Cam. Univ. Press 1957.}

\lref\hardy{G. Hardy and E. Wright, {\it An Introduction to the Theory of Numbers}, Oxford Univ. Press 1979.}

\lref\khinchin{A. Khinchin, {\it Continued Fractions}, Univ. of Chicago Press, 1964.}

\lref\leshouchesvol{See the article by Yoccoz in C. Itzykson, J.-M. Luck, P. Moussa, and 
M. Waldschmidt, eds.   {\it From Number Theory to Physics}, 
Springer Verlag, 1995.}

\lref\cusick{T.W. Cusick and M.E. Flahive, ``The Markoff and Lagrange Spectra,'' Math. Surveys and Monographs, no. 30, 
AMS. 1989.}

\lref\oxtoby{J. Oxtoby, {\it Measure and Category}, Springer-Verlag, GTM vol. 2.}

%\RussoIK
\lref\RussoIK{
J.~G.~Russo and A.~A.~Tseytlin,
``Magnetic flux tube models in superstring theory,''
Nucl.\ Phys.\ B {\bf 461}, 131 (1996)
[arXiv:hep-th/9508068].
%%CITATION = HEP-TH 9508068;%%
}
%\RussoTF
\lref\RussoTF{
J.~G.~Russo and A.~A.~Tseytlin,
``Magnetic backgrounds and tachyonic instabilities in closed superstring
theory and M-theory,''
Nucl.\ Phys.\ B {\bf 611}, 93 (2001)
[arXiv:hep-th/0104238].
%%CITATION = HEP-TH 0104238;%%
}

%\HarveyWM
\lref\HarveyWM{
J.~A.~Harvey, D.~Kutasov, E.~J.~Martinec and G.~Moore,
``Localized tachyons and RG flows,''
arXiv:hep-th/0111154.
%%CITATION = HEP-TH 0111154;%%
}

%\KutasovSV
\lref\KutasovSV{
D.~Kutasov and N.~Seiberg,
``Number Of Degrees Of Freedom, Density Of States And Tachyons In String Theory
%And Cft,''
Nucl.\ Phys.\ B {\bf 358}, 600 (1991).
%%CITATION = NUPHA,B358,600;%%
}

%\KutasovPV
\lref\KutasovPV{
D.~Kutasov,
``Some properties of (non)critical strings,''
arXiv:hep-th/9110041.
%%CITATION = HEP-TH 9110041;%%
}

%\MartinecTZ
\lref\MartinecTZ{
E.~J.~Martinec,
``Defects, decay, and dissipated states,''
arXiv:hep-th/0210231.
%%CITATION = HEP-TH 0210231;%%
}

\lref\hofstadter{D.R. Hofstadter, ``Energy levels and wave functions of 
Bloch electrons in rational and irrational magnetic fields,'' 
Phys. Rev. {\bf B14} (1976) 2239. } 

%\KolVI
\lref\KolVI{
B.~Kol,
``On 6d *gauge* theories with irrational theta angle,''
JHEP {\bf 9911}, 017 (1999)
[arXiv:hep-th/9711017].
%%CITATION = HEP-TH 9711017;%%
}

%\ElitzurPS
\lref\ElitzurPS{
S.~Elitzur, B.~Pioline and E.~Rabinovici,
``On the short-distance structure of irrational non-commutative gauge
%theories,''
JHEP {\bf 0010}, 011 (2000)
[arXiv:hep-th/0009009].
%%CITATION = HEP-TH 0009009;%%
}

%\ChanGS
\lref\ChanGS{
C.~S.~Chan, A.~Hashimoto and H.~Verlinde,
``Duality cascade and oblique phases in non-commutative open string  theory,''
JHEP {\bf 0109}, 034 (2001)
[arXiv:hep-th/0107215].
%%CITATION = HEP-TH 0107215;%%
}

%\CornalbaKD
\lref\CornalbaKD{
L.~Cornalba and M.~S.~Costa,
``Time-dependent orbifolds and string cosmology,''
Fortsch.\ Phys.\  {\bf 52}, 145 (2004)
[arXiv:hep-th/0310099].
%%CITATION = HEP-TH 0310099;%%
}

\lref\slater{N.B. Slater, ``Gaps and steps for the sequence $n\theta\mod 1$,'' Proc. 
Cambridge Philos. Soc. {\bf 63}(1967) 1115-1123.}

\lref\davenport{ H. Davenport, {\it Analytic Methods for Diophantine Equations
and Diophantine Inequalities} (Ann Arbor Publ., 1962), page 13ff.}

\lref\artin{E. Artin, ``Ein mechanisches System mit quasiergodischen 
Bahnen,'' Collected works, p. 499.} 

\lref\series{C. Series,
The modular surface and continued fractions, 
 J. London. Math. Soc. {\bf 31} (1985) 69-80.}

\lref\marklof{
J., Marklof,
``The $n$-point correlations between values of a linear form, 
with an appendix by Z.~Rudnick,''
 Ergod. Th. Dyn. Sys. {\bf 20} (2000) 1127-1172.
}

%\DowkerBT
\lref\DowkerBT{
F.~Dowker, J.~P.~Gauntlett, D.~A.~Kastor and J.~H.~Traschen,
``Pair creation of dilaton black holes,''
Phys.\ Rev.\ D {\bf 49}, 2909 (1994)
[arXiv:hep-th/9309075].
%%CITATION = HEP-TH 9309075;%%
}
%\DowkerUP
\lref\DowkerUP{
F.~Dowker, J.~P.~Gauntlett, S.~B.~Giddings and G.~T.~Horowitz,
``On pair creation of extremal black holes and Kaluza-Klein monopoles,''
Phys.\ Rev.\ D {\bf 50}, 2662 (1994)
[arXiv:hep-th/9312172].
%%CITATION = HEP-TH 9312172;%%
}
%\DowkerGB
\lref\DowkerGB{
F.~Dowker, J.~P.~Gauntlett, G.~W.~Gibbons and G.~T.~Horowitz,
``The Decay of magnetic fields in Kaluza-Klein theory,''
Phys.\ Rev.\ D {\bf 52}, 6929 (1995)
[arXiv:hep-th/9507143].
%%CITATION = HEP-TH 9507143;%%
}
%\DowkerSG
\lref\DowkerSG{
F.~Dowker, J.~P.~Gauntlett, G.~W.~Gibbons and G.~T.~Horowitz,
``Nucleation of $P$-Branes and Fundamental Strings,''
Phys.\ Rev.\ D {\bf 53}, 7115 (1996)
[arXiv:hep-th/9512154].
%%CITATION = HEP-TH 9512154;%%
}

%\CostaNW
\lref\CostaNW{
M.~S.~Costa and M.~Gutperle,
``The Kaluza-Klein Melvin solution in M-theory,''
JHEP {\bf 0103}, 027 (2001)
[arXiv:hep-th/0012072].
%%CITATION = HEP-TH 0012072;%%
}

\def\vt#1#2#3{ {\vartheta[{#1 \atop  #2}](#3\vert \tau)} }

\def\makeblankbox#1#2{\hbox{\lower\dp0\vbox{\hidehrule{#1}{#2}%
   \kern -#1% overlap rules
   \hbox to \wd0{\hidevrule{#1}{#2}%
      \raise\ht0\vbox to #1{}% vrule height
      \lower\dp0\vtop to #1{}% vrule depth
      \hfil\hidevrule{#2}{#1}}%
   \kern-#1\hidehrule{#2}{#1}}}%
}%
\def\hidehrule#1#2{\kern-#1\hrule height#1 depth#2 \kern-#2}%
\def\hidevrule#1#2{\kern-#1{\dimen0=#1\advance\dimen0 by #2\vrule
    width\dimen0}\kern-#2}%
\def\openbox{\ht0=1.2mm \dp0=1.2mm \wd0=2.4mm  \raise 2.75pt
\makeblankbox {.25pt} {.25pt}  }

\def\bun#1/#2{\leavevmode
   \kern.1em \raise .5ex \hbox{\the\scriptfont0 #1}%
   \kern-.1em $/$%
   \kern-.15em \lower .25ex \hbox{\the\scriptfont0 #2}%
}

\def\opensquare{\ht0=3.4mm \dp0=3.4mm \wd0=6.8mm  \raise 2.7pt
\makeblankbox {.25pt} {.25pt}  }

%%%%%%%%%%%%%%%%%%%%%%%

\def\sector#1#2{\ {\scriptstyle #1}\hskip 1mm
\mathop{\opensquare}\limits_{\lower 1mm\hbox{$\scriptstyle#2$}}\hskip 1mm}

\def\tsector#1#2{\ {\scriptstyle #1}\hskip 1mm
\mathop{\opensquare}\limits_{\lower 1mm\hbox{$\scriptstyle#2$}}^\sim\hskip 1mm}
%%%
%%%

\def\frac#1#2{{#1\over#2}}
\def\ZZ{\IZ}
\def\vecnull{{\vec 0}}
\def\CC{{\IC}}
\def\RR{{\IR}}
 
\def\vecp{{\vec p}}
\def\vecm{{\vec m}}
\def\vecx{{\vec  x}}

\def\SLSL{SL(2,\ZZ)\backslash SL(2,\RR)}
\def\SLZ{SL(2,\ZZ)}
\def\SLR{SL(2,\RR)}

\def\scrF{{\CF}}

%-------------------
% title page
%-------------------
%
\Title{\vbox{\baselineskip12pt
\hbox{hep-th/0407150}
\hbox{LPTHE-04-17}
}}
{\vbox{\centerline{Melvin Models and}
\bigskip
\centerline{Diophantine Approximation }}}
\centerline{David Kutasov} 

\smallskip 
\centerline{{\it EFI and Department of Physics}}
\centerline{\it  University of Chicago, Chicago, IL 60637,  USA}

\bigskip
\centerline{Jens Marklof} 

\smallskip 
\centerline{{\it School of Mathematics}}
\centerline{\it  University of Bristol, Bristol BS8 1TW, U.K.}

\bigskip

\centerline{Gregory W. Moore  }
\smallskip 
\centerline{{\it Department of Physics, Rutgers University}}
\centerline{\it Piscataway, NJ 08854-8019, USA}
 \vskip.1in \vskip.1in \centerline{\bf Abstract}  
\noindent
Melvin models with irrational twist parameter provide
an interesting example of conformal field theories with
non-compact target space, and localized states which are
arbitrarily close to being delocalized. We study the 
torus partition sum of these models, focusing on the 
properties of the regularized dimension of the space of 
localized states. We show that its behavior is related to
interesting arithmetic properties of the twist parameter 
$\gamma$, such as the Lyapunov exponent. Moreover, for 
$\gamma$ in a set of measure one the regularized dimension is 
in fact not a well-defined number but must be considered as 
a random variable in a probability distribution.  

%%%***

\Date{July 16, 2004}

%\draftmode

 \def\ap{\alpha'}
\def\vol{\rm vol}
\def\IZ{{\bf Z}}
\def\IC{{\bf C}}
\def\IR{{\bf R}}
\def\mod{{\rm mod}}

\newsec{Introduction}

Two dimensional conformal field theories (CFT's) corresponding
to defects embedded in non-compact target spaces have many 
applications in string theory and are interesting in their 
own right \refs{\MartinecTZ,\HeadrickHZ}.  

As in scattering problems in quantum mechanics, the eigenstates of
the Hamiltonian in such theories split into two classes. 
One consists of delta-function normalizable scattering states,
which can propagate in the whole non-compact space. The other
corresponds to normalizable states localized near the defect. In
order to study the defect, one is particularly interested in the
localized states and their interactions with the scattering states.

An example that has received some attention in recent years
is orbifolds of flat non-compact space. In this case, the 
delocalized (scattering) states belong to the untwisted 
sector of the orbifold, while the localized ones are twisted 
sector states. For orbifolds by a finite group, the spectrum of
localized states is discrete, with finite gaps between states. By
contrast, orbifolds by  infinite groups   can have discrete but
dense spectra of states. The latter case is particularly interesting
since there is then no sharp distinction between localized and
delocalized states.

More generally, while orbifolds by finite groups are well-studied,
orbifolds by infinite groups introduce many new features, and have
not been well-studied (some work has been done on time-dependent 
orbifolds; see \CornalbaKD\ for a review).  
A better understanding of general CFT orbifolds
by infinite groups might provide insights into string cosmology,   the
AdS/CFT correspondence, and noncommutative geometry \MooreFG.

In this note we will study an orbifold of $\IR\times\IC$ by the group $\IZ$,
known as the Melvin model, or the twisted circle. These models 
were introduced and studied in \refs{\DowkerBT\DowkerUP\DowkerGB\DowkerSG\RussoIK-\RussoTF}. For 
further background on
the Melvin model see  \refs{\DavidVM,\HeadrickHZ}
and references therein. We will see that as we vary the orbifold
twist parameter, the model exhibits some unusual behavior,
including divergences associated with a sum
over almost delocalized twisted sector states. These divergences
can be quantified using some results from the theory of Diophantine approximation.
For background on   Diophantine approximation see, e.g.,
\refs{\cassels\hardy\khinchin\schmidt-\leshouchesvol}.

The Melvin CFT is the orbifold
\eqn\model{
(\IR \times \IC)/\IZ}
where the generator $g$, of the group $\IZ$ acts as
\eqn\action{
\eqalign{
y & \to y + 2\pi R \cr
z & \to e^{2\pi i \gamma} z \cr}
}
for $(y,z)\in \IR \times \IC$.
When  $\gamma$ is rational, e.g. $\gamma=1/n$, one can think of the orbifold \action\
as a $\IZ_n$ orbifold of $S^1\times \IC$. For irrational $\gamma$, it is not clear
apriori whether \model\ makes sense as a CFT (and string theory) background.
One of our motivations below will be to explore this issue, by studying the torus
partition sum of the theory. We will see that for irrational $\gamma$ the partition
sum is very sensitive to the number theoretic properties of $\gamma$ 
(physical effects related to the arithmetic of irrational angles have
appeared in some other recent investigations in string theory;
see e.g. \refs{\KolVI,\ElitzurPS,\ChanGS}).

We will mostly focus on the CFT $(\IR \times \IC)/\IZ$.
In string theory on $\IR^{1,6}\times (\IR \times \IC)/\IZ$
the consistency requirements for the existence of the theory
are more stringent, and it is possible
that the theory does not exist for irrational $\gamma$.

Although the orbifold \action\ does not have  fixed points, one
can think of the origin of the $z$-plane as the location of a defect,
near which the twisted states of the orbifold are localized. Indeed,
consider a low lying state in the $w$-twisted sector.  It winds $w$
times around the circle $\IR/2\pi R \IZ$ labelled by $y$. Its
endpoints in the $z$-plane are separated by angle $2\pi \| w \gamma \|$,
where, for a real number $x$, $\| x \|$ denotes the
distance to the nearest integer. \foot{Thus, defining the
fractional part of $x$, $\{ x \} = x-[x]$, one has
$\| x \| = \min(\{ x \}, 1-\{ x \})$.}

A classical string placed a distance $r$ from the origin has energy
\eqn\clssenergy{
\alpha'^2M^2(r) = (R w)^2 + (r \| w \gamma \|)^2~.
}
When $\| w \gamma \|\not=0$, such winding strings 
are localized near the origin -- their wavefunctions fall off 
exponentially as $r\to\infty$. The radial size of such
$w$-twisted strings goes like $1/\| w\gamma\|$. 
In fact, we see from \clssenergy\ that strings stretched in the
angular direction of the $z$-plane behave as if their effective tension
is proportional to $\| w\gamma\|$; this will be
important for our later discussion. After quantization, the reduced
string tension is reflected in the presence of twisted oscillators
for the worldsheet superfield $z$ with moding $\| w\gamma\|$.

For rational $\gamma$, the radial size $1/\| w\gamma\|$
is bounded from above in the twisted sectors. Thus, there
is a clear distinction between localized and delocalized 
sectors. When $\gamma$ is irrational
there are twisted sectors that are arbitrarily close to being delocalized,
since $\| w\gamma\|$ is not bounded from below.

To study the theory for irrational $\gamma$, we would like
to analyze the partition sum of the CFT \model\ on a torus with modulus
$q= e^{2\pi i \tau}$; this corresponds to the trace of
$q^{L_0 - c/24} \bar q^{\bar L_0 - c/24}$ over the eigenmodes
of $(L_0,\bar L_0)$. For non-compact orbifolds, the trace over the
untwisted sector is  divergent -- it is proportional to the volume
of the target space. Sometimes, it is possible to regulate similar volume
divergences by compactifying the space, but here this is not possible without
breaking conformal invariance.

In \HarveyWM\ it was proposed, in a related context, to restrict
the trace in the torus partition sum to the localized states, i.e.
to the twisted sectors of the orbifold. This eliminates the usual
volume divergence from the untwisted sector but, as we will see, leaves
in some cases analogous divergences from ``almost untwisted'' sectors.

The partition sum of the localized states is given by
\eqn\grdd{
Z_{\rm loc}(\tau;\gamma) :=  {\Tr}_{\CH_{\rm loc} } q^{L_0 - c/24} \bar q^{\bar L_0 - c/24}
}
where $\CH_{\rm loc}$ is a sum over twisted sectors
\eqn\deflhol{
\CH_{\rm loc}(\gamma):= \oplus_{ \| w \gamma \|\not=0}\CH_w~.
}
The partition sum $Z_{\rm loc}(\tau;\gamma)$ is not modular invariant.
It transforms under $\tau\to-1/\tau$ to the trace over the untwisted
Hilbert space with a certain projection operator inserted. This is
analogous to what happens for D-branes: the annulus amplitude, which
can be thought of as a trace over open string states whose ends lie on
the D-brane, is related by a modular transformation to a sum over
closed strings that can be emitted by the D-brane.

By analogy to the D-brane case, it was proposed in \HarveyWM\
to study the regularized dimension of the space of localized
states, which is given by \grdd\ in the limit\foot{We set
$\tau_1=0$, such that $q= e^{-2\pi \tau_2}$, and take $\tau_2\to 0$.}
$q\to 1$. For non-compact orbifolds by finite groups
one finds in this limit
\eqn\defgl{
Z_{\rm loc}(\tau\to 0;\gamma)  \sim g_{cl}(\gamma) e^{{\pi c \over 6 \tau_2}}~.
}
The leading exponential term in \defgl\ is universal -- it only
depends on the central charge (or dimension of space). Thus, one
can think of the quantity $g_{cl}$ as a measure of the density of
localized states. Some properties of $g_{cl}$ for finite orbifold
groups were described in \HarveyWM.

As we will see, in the irrational Melvin case 
the coefficient of the exponential in \defgl\ behaves in 
an unusual way and does not have a good limit as $\tau_2\to 0$.
First, it  diverges like $ \tau_2^{-b(\gamma)}$, with some  
constant $b(\gamma)\geq 1/2$.  Moreover - and somewhat surprisingly - 
 the coefficient of 
this divergence, while it is order $1$,  does not have a well-defined 
limit as $\tau_2\to 0$ but varies as a random variable in a probability 
distribution. We explain this point, which is somewhat novel in conformal 
field theory, in sections 3.2, 3.3 below. A rigorous account is given 
in the appendix. One nice aspect of the discussion is that the 
behavior of the regularized dimension is related to the behavior of 
geodesics on a certain modular curve.

The regularized dimension of the space of localized states  \defgl\
is analogous to a similar regularized dimension which proved
useful in RCFT \DijkgraafTF. The D-brane analog of $g_{cl}$
is the product of the tensions of the D-branes on which the
open strings end (see, e.g., \HarveyGQ). Note also that in
models with spacetime fermions, the trace in \grdd\ is usually
taken to include a factor of $(-)^F$, such that spacetime
fermions contribute with a minus sign, and there are usually large
cancellations between bosons and fermions. For the purpose of
estimating the high energy density of states, we should only
sum over spacetime bosons (or over bosons plus fermions); see
e.g. \refs{\KutasovSV,\KutasovPV} for a discussion of the relevant issues.

In the remainder of this note we will study the behavior of the torus
partition sum, and in particular of $g_{cl}$ \defgl, for irrational Melvin
models. The main results are:
\item{(1)} When the twist $\gamma$ of the Melvin model is a Liouville
number of a special kind, the one-loop partition function for bosons
and fermions separately diverges for fixed $\tau$, although the
string theory partition sum $Z_{B} - Z_F $ is finite.  For such twists
it is not clear that the Melvin conformal field theory
makes sense. For $\gamma$ of Diophantine type this pathology is absent.

\item{(2)}  The standard definition \defgl\ of the
regularized dimension determines not a number, but a random variable in 
a probability distribution. This is explained heuristically in sections 3.2 and 3.3. 
A rigorous discussion is given in the appendix. 

\item{(3)} We can use the continued fraction approximations to $\gamma$ to define
a modular invariant regulator in the case of irrational twists. 
We define a degree of delocalization and show that it is
 related to the Lyapunov exponent of $\gamma$.

\newsec{Torus partition sum}

Using the definition of the Melvin CFT \model, \action, one can write the
torus partition sum of the model. In the sector twisted by $g^s$, $s\in \IZ$
($s\not=0$), one has: \foot{Our convention for theta functions is
$$
{\vt{\theta}{\phi}{0} \over  \eta}
= e^{ 2 \pi i \theta \phi}
q^{({\theta^2 \over  2} - {1 \over  24}) }
\prod_{n=1}^\infty (1+ e^{2 \pi i \phi} q^{n-\half + \theta} )
(1+ e^{- 2 \pi i \phi} q^{n-\half -\theta} )
$$
where $\eta$ is the Dedekind eta function.}
\eqn\twistsec{
\eqalign{
&{\Tr}_{\CH_{g^s}} g^t q^{L_0 - c/24}  \bar q^{\bar L_0 - c/24}  = \qquad\qquad \qquad \cr
&{ \vol(\IR) }
\biggl\vert  {\vt{\epsilon_2}{\epsilon_1}{0}\over   \eta^3} \biggr\vert
   \int_{-\infty}^{+\infty} {dp\over 2\pi}  q^{{\ap\over 4}(p+sR/\ap)^2}
 \bar q^{{\ap\over 4}(p-sR/\ap)^2} e^{2\pi i (pR) t}
\Biggl\vert{\vt{\epsilon_2 + s \gamma}{\epsilon_1+t\gamma}{0}\over
\vt{\half + s \gamma}{\half +t\gamma}{0} } \Biggr\vert^2~.\cr}
}
The $|\vartheta/\eta^3|$ prefactor is the contribution of the 
(bosonic and fermionic) oscillators on $\IR$. 
$\epsilon_1,\epsilon_2=0,{1\over2}$ 
label the spin structure of the fermions. The final ratio of theta 
functions is the partition function of the $N=2$ superfield twisted 
by $g^s$ and projected by $g^t$. Note that it only depends on the 
fractional parts $\{ s\gamma\}, \{t\gamma\}$. 

In the orbifold theory we must sum over $g^t$, $t\in \IZ$,
to project onto invariant states, and divide by the order of
the group $\IZ$. We interpret
$$ \vol(\IR)/\vert \IZ \vert=2\pi R~. $$
There is no factor of the
volume of $\IC$ because we are in a twisted sector.
The net result is that the trace in the $g^s$ twisted
sector in the orbifold theory is
\eqn\twistsec{
\eqalign{
&{\Tr}_{\CH_{g^s} }q^{L_0 - c/24}  \bar q^{\bar L_0 - c/24}  = \qquad\qquad \qquad \cr
&2\pi R
\biggl\vert  {\vt{\epsilon_2}{\epsilon_1}{0}\over   \eta^3} \biggr\vert
 \sum_{t\in \IZ}
   \int_{-\infty}^{+\infty} {dp\over 2\pi}  q^{{\ap\over 4}(p+sR/\ap)^2} 
 \bar q^{{\ap\over 4}(p-sR/\ap)^2} e^{2\pi i (pR) t} 
\Biggl\vert{\vt{\epsilon_2 + s \gamma}{\epsilon_1+t\gamma}{0}\over 
\vt{\half + s \gamma}{\half +t\gamma}{0} } \Biggr\vert^2~.\cr}
}
In order to evaluate the $\tau \to 0$ asymptotics it is 
convenient to do the Gaussian integral over $p$ to get
\eqn\zeeloc{
\sqrt{{R^2\over \ap \tau_2} }
\biggl\vert {\vt{\epsilon_2}{\epsilon_1}{0}\over  \eta^3} \biggr\vert
 \sum_{t\in \IZ} e^{-{\pi R^2\over \ap}{\vert t + s \tau\vert^2\over \tau_2}}
\Biggl\vert{\vt{\epsilon_2 + s \gamma}{\epsilon_1+t\gamma}{0}\over
\vt{\half + s \gamma}{\half +t\gamma}{0} } \Biggr\vert^2~.
}
Next we have to sum over the different twisted sectors and spin structures.
The precise details of the sum depend on the particular theory -- CFT on the
orbifold, type 0 or type II string theory on $\IR^{1,6}$ times the orbifold, etc;
see e.g. \mourad\ for a discussion. The different theories behave in a similar
way as far as our analysis is concerned. To be concrete, consider type IIB string
theory on $\IR^{1,1}\times T^5$ times the orbifold. Here, $T^5$ is a five-torus
of volume $V_5$; the compactification is convenient for studying the $\tau_2\to 0$
limit of the partition sum. 

The partition function for the twisted (NS,NS) sectors is
\eqn\bosonicpf{
\eqalign{
& Z_{\rm loc} = {V_5Z_\Gamma\over (2\pi\sqrt{\ap\tau_2 })^5}
\sqrt{{R^2\over \ap \tau_2} }
 \sum_{\| s\gamma\| \not=0,t\in \IZ} e^{-{\pi R^2\over \ap}{\vert t + s \tau\vert^2\over \tau_2}} \cr
&
\Biggl\vert  \half ({\vt{0}{0}{0} \over \eta^3})^3 {\vt{ s \gamma}{ t\gamma}{0}\over
\vt{\half + s \gamma}{\half +t\gamma}{0} }
- e^{-\pi i t s\gamma} \half ({\vt{0}{1/2}{0}\over \eta^3})^3{\vt{ s \gamma}{ 1/2 + t\gamma}{0}\over
\vt{\half + s \gamma}{\half +t\gamma}{0} } \Biggr\vert^2~.\cr}
}
$Z_{\Gamma}$ is a Siegel-Narain theta function of signature $(5,5)$
corresponding to $T^5$.  The behavior of the partition sum in
the limit \defgl\ does not depend on the details of the compactification.

To analyze the $\tau\to 0$  asymptotics of  \zeeloc\ we need the following
asymptotics for $\tau=i\beta \to 0$, with $\beta$ real
\eqn\etaasym{
\vert \eta(\tau)\vert \to {1\over  \sqrt{ \beta}}  e^{-{2\pi \over 24  \beta}}
}
Similarly, 
\eqn\thetastoi{
\vt{\theta}{\phi}{0} \rightarrow  
\cases{
\beta^{-1/2} e^{2\pi i \theta} \tilde q^{\half \| \phi\|^2}, & if $\half < \phi < 1$ \cr
\beta^{-1/2} (1+e^{2\pi i \theta}) \tilde q^{\half \| \phi\|^2}, & if $  \phi =\half  $ \cr
\beta^{-1/2}  \tilde q^{\half \| \phi\|^2}, & if $0\leq   \phi < \half  $ \cr}
}
where $\tilde q = \exp(-2\pi/\beta)$. 
Using these asymptotic formulae one can check that the leading behavior
arises from $\epsilon_1=t=0$ in \zeeloc, \bosonicpf. One finds that
\eqn\remainder{
Z_{\rm loc} \quad  \sim \quad {1 \over 16} \sqrt{{R^2\tau_2 \over \ap } }e^{2\pi\over \tau_2}
 \sum_{\| s\gamma\| \not=0}^\infty e^{-{\pi R^2 \tau_2 \over \ap}   s^2}
{1\over ( \sin\pi s \gamma)^2}~.
}
Before discussing the mathematical properties of \remainder\
let us interpret the crucial factor $1/(\sin\pi s \gamma)^2$
in \remainder. As mentioned in the discussion following
\clssenergy, twisted sectors with $\| s\gamma\|<<1$
give rise to nearly delocalized states whose radial size scales like
$1/\| s\gamma\|$. This is reflected in the spectrum
of $L_0$ in the following way. In the $s$-twisted sector,
all states wind $s$ times around the $y$ circle, and thus
have a large (for large $s$) ground state energy, of order
$Rs$ (or $L_0\sim (Rs)^2$, see \clssenergy). This gives the exponential 
prefactor in the sum \remainder. On top of this ground state energy, when
$\| s\gamma\|$ is small, one finds a narrowly-spaced spectrum
of states, associated with the twisted oscillators of the superfield 
$z$. This gives the inverse sine factor in \remainder. Thus, we 
see that this factor is directly related to the spatial extent of  
the twisted states.

If $\gamma $ is rational, $\gamma = p/q$ in lowest terms,\foot{Here
$q$ is an integer, not to be confused with the modular parameter
$q=e^{-2\pi \tau_2}$. } \remainder\ has
a smooth $\tau_2\to 0$ limit.  The limit is  easily evaluated by  
setting $s= q\ell + j$, $0 \leq j< q-1, \ell \in \IZ$ to get 

\eqn\remainderp{
Z_{\rm loc} \quad \sim \quad e^{2\pi/\tau_2} {1\over 16}   
\sqrt{{R^2\tau_2 \over \ap } } \sum_{j=1}^{q-1} 
\sum_{\ell \in \IZ }e^{-{\pi R^2 q^2 \tau_2 \over \ap}   
(\ell + j/q)^2} {1\over (2 \sin\pi p j/q)^2}~.
}
Taking the $\tau_2\to 0$ limit we reproduce the familiar
expression for the $\IC/\IZ_q$ orbifold   \HarveyWM: 
\eqn\remainderp{
 \dim\CH_{\rm loc}(\gamma = p/q) =  {1\over 16 q} \sum_{j=1}^{q-1} 
{1\over (\sin\pi p j/q)^2}~.
}
The trigonometric sum is easily evaluated \HarveyWM, 
\eqn\gclrat{
\dim\CH_{\rm loc}(\gamma = p/q) = {1\over 48} (q-{1\over q}) ~.
}
We see that for rational $\gamma$, the Melvin model
is closely related to the corresponding $\IC/\IZ_q$ orbifold.
Note that the result only depends on $q$ and hence is a highly 
erratic function of $\gamma\in {\bf Q}$. This is the first
indication that we are dealing with delicate functions of $\gamma$.

\newsec{Comments on the sum in the case of irrational $\gamma$ }

Now we turn to the case of $\gamma$ irrational. Stripping the universal
exponential in \defgl\ from  \remainder, we see that to compute $g_{cl}$
we need to evaluate
\eqn\sumtry{
g(y;\gamma):=
\sqrt{y}
\sum_{s\not= 0}^\infty e^{-\pi y s^2} {1\over \sin^2 \pi s \gamma}
}
in the limit $y\to 0$ (here $y = \tau_2 R^2/\ap$).

First, note that it is not obvious that the sum converges
for finite $y$ (or $\tau_2$). Indeed, we will see in section 4 that for certain
transcendental numbers it diverges. However, for a ``large'' class of irrational
numbers, including all algebraic numbers,  it does  converge. Recall the standard:

\noindent
{\bf Definition:} An irrational number is of Diophantine type $(K,\sigma)$ if
for all $q\geq 1$, 
\eqn\diophantype{
\sigma_q(\gamma):={\rm inf}_{1\leq s \leq q} \| s \gamma\| \geq {K\over q^{1+\sigma} }~.
}
We denote the set of numbers of Diophantine type $(K,\sigma)$ by 
$\CD(K,\sigma)$, and we also denote 
\eqn\diotypeii{
\CD(\sigma) := \cup_{K>0} \CD(K,\sigma)~.
}
If $\gamma$ is of Diophantine type $(K,\sigma)$ then
$g(y;\gamma)$ exists for all positive $y$. To show this 
we use  %
\eqn\sninequs{
2 \| z \| \le|\sin \pi z| < \pi \| z \|~,
}
(the best estimate valid for all real, non-integer $z$)
to put upper and lower bounds on $g(y;\gamma)$:
\eqn\sumappx{
\sum_{s\not=0} {1\over \pi^2 \| s\gamma\|^2} e^{- \pi y s^2} < 
\sum_{s\not=0}  e^{- \pi y s^2} {1\over \sin^2 \pi s \gamma}<  
\sum_{s\not=0} {1\over 4 \| s\gamma\|^2}  e^{-\pi y s^2}~.
}
If $\gamma$ is of type $(K,\sigma)$ then 
\eqn\dioest{
{1\over \| s\gamma\|} \leq s^{1+\sigma}/K
}
and hence by \sumappx\ the series is bounded above by a convergent sum.

Some interesting  facts, which can be found in
\refs{\cassels\hardy\khinchin-\leshouchesvol} are,
first, that the set   $\CD(\sigma)$ is invariant under $SL(2,\IZ)$
(acting via fractional linear transformations on the elements of
 $\CD(\sigma)$). Second, a theorem of Roth says that if $\gamma$
 is algebraic of degree $\geq 2$   then it is of type $(K,\sigma)$ for
all $\sigma >0$ and some $K$.
\foot{A much easier theorem of Liouville, which is all we need
to establish convergence for algebraic numbers, says that a degree $n\geq 2$
algebraic number is of Diophantine type $(K, n-2)$. }

Diophantine approximation can give  us some idea
of what the asymptotics of $g(y;\gamma)$
might be like. If there are many very good
rational approximants to $\gamma$ then
$\sin \pi s \gamma$ is ``often'' close to
zero, and we expect a divergence as $y\to 0$.
If good rational approximants to $\gamma$ are ``rare''
then the lower limit in \sumappx\ is more accurate
and $g(y;\gamma)$ will grow more slowly.

What we can say rigorously is that  if $\gamma$ is of Diophantine
type $(K,\sigma)$ then, from \dioest\
\eqn\rigr{
C_1\leq g(y;\gamma)\leq C_2y^{-\sigma-1}
}
for some constants $C_i$.
Therefore, we can define a non-negative number $b(\gamma)$ by :
\eqn\rigrii{
b(\gamma):= {\rm inf} \{b: \lim_{y\to 0} y^b g(y;\gamma)=0 \}~.
}
We next show that $b(\gamma)\geq 1/2$.

\subsec{A lower bound for $b(\gamma)$ }

To show that $g(y;\gamma)$ always diverges for $y\to 0$ 
at least as strongly as $1/\sqrt y$, we use the continued
fraction expansion in positive integers $a_n$:
\eqn\contfrc{
\gamma = [a_0, a_1, a_2, \dots ] =a_0 + {1\over a_1 + {1\over a_2 + \cdots}~.}
}
The integers $a_n$ are known as {\it partial quotients}. 
The best rational approximants to $\gamma$ are always 
provided by the {\it convergents}
\eqn\convergnets{
{p_n\over q_n} := [a_0,\dots, a_n] 
}
 in the continued fraction expansion: 
\eqn\convergents{
\vert \gamma - {p_n \over q_n} \vert < {1\over q_n^2}~.
}
The $q_n$ grow exponentially as a function of $n$. 
Roughly speaking,
\eqn\libup{
q_n \sim c e^{\half \lambda(\gamma)n }
}
and more rigorously:
\foot{There are $\gamma$'s for which the limit does not exist.}

\eqn\liups{
\lambda(\gamma):= 2 \lim_{n\to \infty} {1\over n} \log q_n~.
}
The quantity $\lambda(\gamma)$ is known as the Lyapunov exponent of $\gamma$.

Taking a lower bound on $g(y;\gamma)$ by summing only over $s=q_n$ 
and using \sumappx, \convergents, one can show that
\eqn\lowbund{
g(y;\gamma) > 
{2\over \pi^2} \sqrt{y} \sum_{n=1}^\infty {e^{-\pi y q_n^2} 
\over \| q_n \gamma \|^2} > 
{2\over \pi^2} \sqrt{y} \sum_{n=1}^\infty q_n^2 e^{-\pi y q_n^2}~.
} 
Now, using \libup\ we see that   the divergence as $y\to 0$ is at least as strong as
\eqn\lowerbddv{
g(y;\gamma)\geq {1\over y^{1/2} } {2 \over \pi^3 \lambda(\gamma)}~.
}

\subsec{$g(y;\gamma)$ and the three gap theorem}

Some further insight can be gained on the behavior of 
$g(y;\gamma)$ as $y\to 0$ using the three-gap theorem 
of \slater. 

The asymptotics of $g(y;\gamma)$ as $y\to 0$ 
are the same as the $N \to \infty $ asymptotics of  the sum
\eqn\geen{
g_N(\gamma)=N^{-1} \sum_{n=1}^N \| n \gamma \|^{-b}
}
in the case $b=2$, where we identify $y \sim 1/N^2$.  It is
useful in the discussion to keep $b$ general.

In the case $b\leq 0$, Kronecker's theorem, which
tells us that $\| n \gamma \|$ are uniformly distributed
implies
\eqn\kronecker{
g_N(\gamma) \to \int_0^1 \| x \|^{-b} dx.
}
The same holds when $0<b<1$, however one needs to assume $\gamma$ is
Diophantine of type $\sigma$ where $\sigma$ depends on $b$. This is because
values close to zero might cause some divergence since $\| x \|^{-b}$ is unbounded
there. Estimates for the case $b=1$ are also classic \davenport.

We are here interested in $b>1$. In this case
the sum is dominated by a finite number of terms.
In order to see this  order the points $\| n \gamma\| $ ($n=1,\ldots,N$) in the
interval $[0,1/2]$ and label them by
\eqn\insti{
0<\xi_1 < \ldots < \xi_N <1/2.
}
So
\eqn\insti{
g_N(\gamma)=N^{-1} \sum_{n=1}^N \xi_n^{-b} .
}

We now summarize the results of \slater. Label the fractional parts of $n\gamma$   by
\eqn\insti{
0<\eta_1 < \ldots < \eta_N < 1 .
}
The ``three gap theorem'' states that every  spacing $\eta_{n+1}-\eta_n$
is equal to either $\alpha$, $\beta$ or $\alpha+\beta$, 
where $\alpha=\eta_1$ and $\beta=1-\eta_N$. In \slater\ one finds  
formulae for $\alpha,\beta$ in terms of the continued fraction approximation of
$\gamma$. In particular $\alpha$ and $\beta$ in general have no asymptotics
as $N\to\infty$. 

Now if $\gamma$ is of bounded
type (i.e. if   the partial quotients $a_n$ are bounded by some constant), 
one finds immediately from the three gap theorem 
that there are constants $c,C>0$ such that
\eqn\insti{
c/N \leq \alpha \leq C/N, \qquad
c/N \leq \beta \leq C/N .
}
It is thus natural to write
\eqn\insti{
g_N(\gamma)=N^{b-1} \sum_{n=1}^N (N\xi_n)^{-b} .
}
Since the gaps between the $N\eta_n$ are bounded from below by a constant,
we have $c'n\leq N\xi_n\leq C' n$ for suitable constants $c',C'>0$.
Therefore (and provided $b>1$),
given any error threshold $\epsilon>0$ we find an $M_\epsilon$ so that
\eqn\insti{
\limsup_{N\to\infty} \sum_{n=M_\epsilon}^N (N\xi_n)^{-b} < \epsilon.
}
Hence
\eqn\insti{
g_N(\gamma)=N^{b-1} \sum_{n=1}^{M_\epsilon-1} (N\xi_n)^{-b} + 
O(\epsilon N^{b-1}).
}
This means $g_N(\gamma)$ is of order $N^{b-1}$,
and furthermore arbitrarily well approximable 
by a finite number of terms. Recall the $N\xi_n$, $n=1,\ldots,M_\epsilon$,
are bounded from above and below and
have an explicit expression in terms of the continued fraction approximants
of $\gamma$.

One important consequence of these considerations is that $\sqrt{y}g(y;\gamma)$ has 
no good asymptotics. The value remainds bounded but fluctuates as $y \to 0$. 
Nevertheless, as we will see in the next section, this value is governed by a 
definite probability law.

\subsec{The regularized dimension is a random variable in a probability distribution}

We have seen in the previous subsection that the asymptotic $y\to 0$ behaviour
of the function  
\eqn\markli{
g(y;\gamma)= 
\sqrt y \sum_{m\in\ZZ-\{0\}} \frac{e^{-\pi m^2 y}}{\sin^2(\pi m\gamma)}
}
is determined by the continued fraction expansion of $\gamma$. 
We will here refine our analysis by exploiting the dynamical properties
of the geodesic flow on the modular surface. 
The connection between continued fraction dynamics and geodesic flow
is non-trivial but well understood, cf. \refs{\artin,\series}.

To explain the strategy, note  that
\eqn\marklii{
\tilde g(y;\gamma)=  \sqrt y \sum_{(m,n)\in\ZZ^2-\{\vecnull\}}
\frac{e^{-\pi y m^2}}{\pi^2(m \gamma +n)^2} 
}
has the same asymptotic behaviour as $g(y;\gamma)$, up to an error of
order $O(1)$, i.e.,
\eqn\markli{
g(y;\gamma)=\tilde g(y;\gamma)+O(1),
}
uniformly for all $\gamma$. 
(To prove this use the identity 
$$
{1\over \sin^2(\pi m \gamma)} = {1\over \pi^2} \sum_{n=-\infty}^{+\infty} {1\over 
(m \gamma + n)^2} 
$$
and add and subtract the $(m=0,n\not=0)$ terms by hand.) 
The main idea is now to construct
a certain modular function $F(M)$ on $\SLSL$,
such that
\eqn\markli{
\sqrt y \tilde g(y;\gamma) = F(M(t)), \qquad t=-\log y\to\infty,
}
where $M(t)\in\SLR$ is evaluated along the geodesic
\eqn\markli{
M(t)=\pmatrix{ 1 & \gamma \cr 0 & 1 \cr}
 \pmatrix{ e^{-t/2} & 0 \cr 0 & e^{t/2} \cr},
\qquad t\geq 0.
}
The asymptotics of $\sqrt y \tilde g(y;\gamma)$ is now entirely
determined by the geometric distribution of the geodesic associated
with a particular value of $\gamma$. For example:

(a) If $\gamma$ is a quadratic irrational, then the geodesic $M(t)$
is asymptotic to a closed geodesic with period $T_\gamma$. Hence
\eqn\markli{
e^{-t/2} \tilde g(e^{-t};\gamma) \sim \phi(t)
}
where $\phi(t)$ is a bounded periodic function with period $T_\gamma$. 

(b) If $\gamma$ is badly approximable by rationals (i.e., Diophantine of
bounded type), then the geodesic $M(t)$
is asymptotic to a geodesic which never leaves a bounded set in $\SLSL$. 
Hence $e^{-t/2} \tilde g(e^{-t};\gamma)$ is
bounded for all $t$. Our analysis will show that
in general $F(M)$ is a non-constant function, hence 
$e^{-t/2} \tilde g(e^{-t};\gamma)$ does not converge to a constant.

(c) For almost all $\gamma$ (with respect to Lebesgue measure) the 
corresponding geodesic $M(t)$ becomes equidistributed in $\SLSL$,
a consequence of the ergodicity of the geodesic flow. Hence the 
fluctuations of $e^{-t/2} \tilde g(e^{-t};\gamma)$ on some
long stretch $[0,T]$ ($T\to\infty$)
have the same probability distribution as the function $F(M)$,
where $M$ varies over $\SLSL$. That is,
\eqn\probdist{
\frac{1}{T} \int_0^T \delta(X-e^{-t/2}
g(e^{-t};\gamma)) dt \longrightarrow P(X)= \int_{\SLSL} \delta(X-F(M))dM.
}
Interestingly, the limit distribution has an algebraic
tail, $P(X)\sim A X^{-3/2}$, and hence no first moment.
See Theorem 1 in the appendix  for details.

\newsec{Divergence for finite $\tau$}

In the previous section we discussed the behavior of the
torus partition sum \bosonicpf\ in the limit $\tau\to0$. 
In this section we will see that for some $\gamma$, the
sum over twisted sectors (and thus \sumtry) diverge for
finite $\tau$. This point has been mentioned briefly in \LiuKB. 

The dangerous factor in the partition sum \bosonicpf\
is the function 
\eqn\baddenom{
\vt{\half + s \gamma}{\half  }{0}
}
which appears in the denominator; it becomes very small when 
$\| s\gamma\|<<1$. As we have seen, this is due 
to the fact that the corresponding states are nearly delocalized.

Consider
\eqn\smallfun{
F(x) := \vt{\half + x}{\half  }{0} ~. 
}
It is easy to check that $F(x+1)=F(x)$ and $e^{-i \pi x} F(x)$ is an 
odd function of $x$ given at small $x$ by 
\eqn\smallex{
e^{-i \pi x}F(x) = -(2\pi\tau \eta^3 )x + + (2\pi^3 \tau^3 E_2 - 12\pi^2 i \tau^2)\eta^3 {x^3\over 3!} + \cdots
}
The convergence of the sum over twisted sectors
of \bosonicpf\ for fixed $\tau$ is controlled by
\eqn\estimate{
\sum_{s=1}^\infty {1\over \| s \gamma\|^2}
e^{-{\pi R^2\over \ap}s^2 {\vert  \tau\vert^2\over \tau_2}}
}
As discussed in the previous section,
for $\gamma$ of Diophantine type 
$(K,\sigma)$ the sum \estimate\ converges. 
On the other hand, for certain Liouville 
numbers the sum actually diverges. To show 
this, consider the subsum given by $s=q_n$ 
where $q_n$ is the denominator of the 
convergents of $\gamma$, \convergnets. Then \schmidt:
\eqn\bounds{
{1\over q_n + q_{n+1}  } < \| q_n \gamma \| < {1\over q_{n+1}}~.
}
Thus, \estimate\ is bounded from below by
\eqn\lowbd{
\sum_{n} q_{n+1}^2 e^{- \kappa  q_n^2}
}
where $\kappa $ is some constant.
Now if
\eqn\sthrj{
2 \log q_{n+1} - \kappa  q_n^2 = \CO(1)
}
or is even bounded below by $- \log n$ then the series \lowbd\ diverges.

We can thus construct numbers for which the  series \lowbd\  diverges by considering
$\gamma$ of the form
\eqn\transgamma{
\gamma = \sum_{n=1}^\infty {1\over 10^{f(n)}}
}
for certain rapidly increasing functions $f(n)$. Indeed  we may take the
subsum with $s=10^{f(n)}$. Then
\eqn\trnsgmmma{
\vert \gamma - \sum^{n}_{j=1} {1\over 10^{f(j)}} \vert < {2\over 10^{f(n+1)}}~.
}
Now consider  any function $f(n)$ that satisfies
an equation of the form
\eqn\recursion{
f(n+1) = f(n) + \kappa  10^{2 f(n)} + g(n)
}
where $g(n)$ is, say, any positive function of $n$. Then, using $q_n = 10^{f(n)}$ and 
$q_{n+1}^2 < \| q_n \gamma\|^{-2}$ we see that for such functions $f(n)$ the series 
\estimate\ diverges. Thus, there are
continuum many transcendental numbers for which the sum diverges.

\newsec{An alternative regularization using continued fractions}

In the previous sections we discussed the regularized number of
localized states given by the partition sum \grdd\ in the limit
$\tau\to 0$. We saw that there are some irrational numbers 
for which the sum over localized states diverges even for finite 
$\tau$. This divergence is due to the effect of ``nearly untwisted strings''
with $\| w\gamma\|$ small. It is natural to ask whether
one can regularize this divergence in some other way, consistent 
with conformal symmetry and modular invariance.

Replacing $\IC$ by, say, a sphere of finite radius breaks conformal
symmetry, and introduces subtle questions of orders of limits. 
Similarly, putting a cutoff on the sum over twist sectors breaks
modular invariance. One simple way to regulate the volume divergence
is to use the continued fraction expansion of $\gamma$
\eqn\contdfr{
\gamma = [0, a_1, a_2, \dots ]~.
}
Cutting off the continued fraction at a finite
place leads to the rational convergents:
\eqn\gammsf{
\gamma^{(n)} := [0,a_1, a_2, \dots, a_n] := {p_n\over q_n}~.
}
For the rational twists  $\gamma^{(n)}$ we have a clear 
separation of  localized from delocalized states and the 
regularized dimension of the space of localized  states is \gclrat
\eqn\regshd{
\dim \CH_{\rm loc}(\gamma^{(n)}) = {1\over 48} (q_n - 
{1\over q_n})~.
}
Similarly, other correlation functions in the orbifold CFT 
are well-defined for finite $n$.

One can formally think of the original orbifold with twist
parameter $\gamma$ as the limit $n\to \infty$ of \gammsf.
Of course, $q_n\to \infty $ as $n\to \infty$, but it does 
so at different rates for different $\gamma$'s; the rate
depends sensitively on $\gamma$ through the Lyapunov 
exponent \liups.

The exponential growth of $q_n$ suggests that we should define an
``entropy of delocalization''  by considering the limiting behavior 
of 
$S_n(\gamma) = \log \dim \CH_{\rm loc}(\gamma^{(n)} )$.
With this measure of delocalization we have
\eqn\delcoa{
{\log \dim \CH_{\rm loc}(\gamma_1) \over \log \dim \CH_{\rm loc}(\gamma_2)}=
\lim_{n\to \infty}  {S_n(\gamma_1) \over S_n(\gamma_2) } = {\lambda(\gamma_1) \over \lambda(\gamma_2) }~.
}

Some interesting facts about $\lambda(\gamma)$, which can be found in  \pollicott, are
the following. First, for almost every $\gamma$, $\lambda(\gamma)$ is given by Khinchin's constant
\eqn\kinsh{
\lambda_0 = {\pi^2 \over 6 \log 2}~.
}
Moreover, the  range of $\lambda(\gamma)$ as $\gamma$ runs over irrational numbers in
$(0,1)$ is
\eqn\rangegamm{
[2\log{1+ \sqrt{5} \over 2}, \infty)~.
}
Thus, the entropy of delocalization is a nontrivial function 
of the twist parameter $\gamma$ of the Melvin model.

\bigskip
{\bf Remarks}

\item{1.} One very interesting property of the Lyapunov 
exponent  $\lambda(\gamma)$ is that it is invariant under 
$SL(2,\IZ)$ acting on $\gamma$ via fractional linear transformations.
This is easily seen since 
$\gamma \to \gamma +1$ obviously does not change the 
exponent while, for $\gamma = [0,a_1, a_2, \dots]$\ 
we have $1/\gamma = [a_1 ,a_2, a_3, \dots]$, so 
$\{1/\gamma\} = [0,a_2, a_3, \dots]$. 

\item{2.} The Lyapunov exponent of $\gamma$ is indeed a 
Lyapunov exponent for a dynamical system, namely that 
defined by the Gauss map   $T(x) = \{ 1/x \}$, which 
shifts the entries of the continued fraction expansion. 
In this context it is quite amusing to note that a naive 
analysis of the GLSM description for the Melvin model 
discussed in \DavidVM\  appears to lead to a connection
between 2D RG flow and the Gauss map. If we choose a  
GLSM  with gauge group $\IR$ with gauge group action
$(X_1, X_2, P) \to (e^{i \gamma \theta } X_1, e^{-i \theta} X_2, P + i \theta)$ 
then the standard analysis of the $D$-term equation
\eqn\dterm{
\gamma \vert X_1 \vert^2 - \vert X_2 \vert^2 = p_1
}
suggests that the Melvin geometry with twist parameter 
$\gamma$ and radius $R$ flows to that with twist parameter 
$\{1/\gamma\}$ and radius $R/\gamma$. Thus, at least as long
as $R$ remains small, the flow from the UV to the IR acts as
a Gauss map on $\gamma$.
\item{3.} Some other interesting relations of Lyapunov exponents 
to areas of physics are explored in \marcolli.
\item{4.} The approach of this section has the advantage 
that it can be easily extended to other twisted tori geometries
$\IC^d \times \IR^{d'}/\Gamma$, where $\Gamma$ acts by linear
transformations in $\IC^d$ and by translations in $\IR^{d'}$.

\newsec{Melvin models in string theory}

So far we have been focusing on the conformal field 
theory of the Melvin orbifold, and the divergences 
associated with the sums over twisted sectors in 
defining partition functions in this CFT. In string 
theory we have the further complication that we must 
integrate amplitudes over moduli space.

We have seen that for certain irrational numbers $\gamma$,
the partition function of the twisted (NS,NS) sectors,
which contain spacetime bosons, is divergent for
fixed $\tau$. In type II string theory, what enters into
the torus amplitude -- the one loop contribution to the
cosmological constant -- is the difference of spacetime
bosons and fermions,
\eqn\zeeloc{
\sqrt{{R^2\over \ap \tau_2} }
 \sum_{(s,t)\not= (0,0)} e^{-{\pi R^2\over \ap}{\vert t + s \tau\vert^2\over \tau_2}}
\Biggl\vert{ \sum_{\epsilon_1, \epsilon_2} \eta_{\epsilon_1,\epsilon_2}
({\vt{\epsilon_2}{\epsilon_1}{0}\over  \eta})^3
\vt{\epsilon_2 + s \gamma}{\epsilon_1+t\gamma}{0}\over
\vt{\half + s \gamma}{\half +t\gamma}{0} } \Biggr\vert^2
}
where $\eta_{\epsilon_1\epsilon_2}=\pm 1$ are described in \mourad.
It is still the case that the denominator of \zeeloc\ is small
in sectors with $t=0$ and small $\| s\gamma\|$ \baddenom. 
However, the numerator goes to zero as well, due to the standard Riemann 
identity, or more physically because when $\| s\gamma\|\to 0$, 
one is approaching the untwisted contribution, which vanishes due to
the standard supersymmetric cancellations. However, when $t\gamma$ is 
a good approximation to an {\it odd} integer the numerator does not 
cancel the denominator.
\foot{We thank the referee for pointing out this subtlety.} 
In this case there are potential divergences 
such as those discussed in section 4.

In addition to this, for other one loop amplitudes we expect that 
the sum over twisted sectors $s$ will again be problematic. 
This is compounded by the fact that we must integrate over 
moduli space, since the integral over $\tau_2$ has the form 
\eqn\ofthrfm{
\int^{\infty} d\tau_2 \tau_2^\nu  e^{- s^2 \tau_2} \sim {1\over s^{(\nu+1)/2} }~.
}
Thus the exponential suppression in $s$ is replaced by power law suppression, 
which can  be easily overwhelmed by $1/\| s\gamma\|^2$
even for algebraic irrationals $\gamma$. For this reason, it is far from 
clear to us that Melvin spacetimes with irrational values of $\gamma$ are 
well-defined string backgrounds. This issue merits further investigation.

\newsec{Discussion} 

The discussion of the previous sections
suggests that physics is not continuous 
as a function of $\gamma$. This may seem 
physically unreasonable. How can continuous 
changes of a magnetic field lead to discontinuous 
conformal field theory or string theory amplitudes? 
There is a well-known precedent for this kind of
behavior, namely the Azbel-Hofstadter model of an 
electron in a magnetic field in the presence of 
a periodic potential \hofstadter.  The spectrum
of the Schr\"odinger operator is a sensitive function 
of the magnetic field, and depends on its arithmetic 
nature.

It has been claimed that Melvin models provide a 
smooth interpolation between IIB and 0A string 
theory and this has been used to argue that the 
endpoint of 0A tachyon condensation is the IIB theory
\refs{\CostaNW,\GutperleMB,\DavidVM}. It is possible that 
the interpolating string theories with irrational 
$\gamma$ do not exist, thus calling these claims into 
question.

A natural class of questions which arise in the context of 
our considerations have to do with asking
``how many'' $\gamma$'s  or ``how often'' a $\gamma$ leads 
to a divergent model, or to a model with fixed Lyapunov 
exponent, and so forth.  These are subtle and difficult 
questions. They  have been the subject of much research.  
To quote one nice result \pollicott, the set of $\gamma$ 
such that $\lambda(\gamma)$ takes a value larger than the 
Khinchin constant $\lambda_0$ \kinsh\ has positive Hausdorff 
dimension. Further discussion of such matters would take us 
into the subject of measure and category \oxtoby, so this 
seems a good place to stop.

\bigskip  
\noindent{\bf Acknowledgements:}  
We would like to thank J. Harvey and E. Martinec 
for collaboration on related matters and for
useful discussions.  We  would like to thank J. Lagarias, 
J. Maldacena,   and B. Pioline  for useful discussions and 
correspondence.  We also thank the
LPTHE at Jussieu, Paris, for hospitality while this 
paper was written.  The work of GM is
supported in part by DOE grant DE-FG02-96ER40949.  That
of DK is supported in part by DOE grant DE-FG02-90ER40560.
JM was supported by an EPSRC Advanced Research Fellowship and
the EC Research Training Network (Mathematical Aspects of Quantum 
Chaos) HPRN-CT-2000-00103.

\appendix{A}{Detailed proof of \probdist }

\subsec{Sums over lattice points}

For any $M\in\SLR$ consider the sum
\eqn\proofi{
F(M)=\sum_{\vecm\in\ZZ^2-\{\vecnull\}} f(\vecm M)
}
where $\vecm$ runs over all non-zero integer row vectors, 
and $f$ is a positive function on $\RR^2$.
Since the modular group $\SLZ$ leaves the lattice $\ZZ^2$ and the origin
$\vecnull$ invariant,
we have immediately
\eqn\proofi{
F(KM)=F(M)
}
for every $K\in\SLZ$. $F$ may thus be viewed as a function
on the homogeneous space $\SLSL$. There is a simple formula for the
average of $F$ with respect to Haar measure $dM$, normalized as a probability
measure so that
\eqn\proofi{
\int_{\SLSL} dM =1.
}
We then have
\eqn\proofi{
\int_{\SLSL} F(M) dM = \int_{\RR^2} f(\vecx) dx . 
}
This is a special case of Siegel's weight formula for $SL(d,\RR)$; 
for a proof see e.g.  Theorem 3.15 of \marklof.

The function $\tilde g(y;\gamma)$
is connected to an automorphic function $F$ of the above form:
choose 
\eqn\proofi{
f(x_1,x_2)= \frac{e^{-\pi x_1^2} }{\pi^2 x_2^2} .
}
Then, at the point
\eqn\proofi{
M=\pmatrix{ 1 & x \cr 0 & 1 \cr}
 \pmatrix{ y^{1/2} & 0 \cr 0 & y^{-1/2} \cr},
}
we have
\eqn\proofi{
F(M)=\sqrt{y}\, \tilde g(y;x).
}
The space $\SLSL$ can be identified with the unit tangent bundle of the
modular surface by means of the Iwasawa decomposition
\eqn\proofi{
M=\pmatrix{ 1 & x \cr 0 & 1 \cr}
 \pmatrix{ y^{1/2} & 0 \cr 0 & y^{-1/2} \cr}
 \pmatrix{ \cos(\theta/2) & \sin(\theta/2) \cr
-\sin(\theta/2) & \cos(\theta/2) \cr};
}
$z=x+i y$ are the standard upper half plane coordinates and the angle
$\theta\in[0,2\pi)$
describes the direction of the unit tangent vector at $z$.
This identification induces the following action of 
a matrix $\pmatrix{ a & b \cr c & d \cr}\in\SLR$
on a point $M=(z,\theta)$:
\eqn\proofi{
\pmatrix{ a & b \cr c & d \cr} (z,\theta)
= \left( \frac{az+b}{cz+d}, \theta - 2\arg(cz+d) \right).
}
A fundamental domain of $\SLZ$ in these coordinates is
\eqn\proofi{
\scrF=\{ (z,\theta): |z|>1, |x|<1/2, \theta\in[0,2\pi)\}.
}
The normalized Haar measure reads
\eqn\proofi{
dM = \frac{3}{2\pi^2} \, \frac{dx\,dy\,d\theta}{y^2}.
}

The geodesic flow on $\SLSL$ is represented by the right translation
\eqn\proofi{
M(0) \mapsto M(t)=M(0) \Phi^t, \qquad \Phi^t = 
\pmatrix{ e^{-t/2} & 0 \cr 0 & e^{t/2} \cr}.
}
The values of $\tilde g(y;x)$ are thus those of
$F(M)$ evaluated along a geodesic $M=M(t)$ with initial 
condition $M(0)=\pmatrix{ 1 & x \cr 0 & 1 \cr}$.

\subsec{Singularities}

To analyze the singularities of $F(M)$ we split
\eqn\proofi{
F(M)=F_0(M)+F_1(M)
}
where $F_0,F_1$ are defined in the same way as $F$ above
with $f$ replaced with
\eqn\proofi{
f_0(x_1,x_2)= \frac{e^{-\pi x_1^2} }{\pi^2 x_2^2} \chi_0(x_2)
}
and
\eqn\proofi{
f_1(x_1,x_2)= \frac{e^{-\pi x_1^2} }{\pi^2 x_2^2} \chi_1(x_2),
}
respectively. $\chi_0$ and $\chi_1$ are continuous functions with
values in $[0,1]$ such that $\chi_0(x)+\chi_1(x)=1$, and 
\eqn\proofi{
\chi_1(x)=
\cases{ 
1, & $x\in[-\epsilon,\epsilon]$\cr
0, & $x\notin[-\epsilon(1+\epsilon),\epsilon(1+\epsilon)]$\cr
}
}
for some fixed $\epsilon>0$. (The extra $(1+\epsilon)$ factor is used
to accommodate the continuity of $\chi_1$; we think of 
$\chi_0$ and $\chi_1$ as smoothed
characteristic functions.)

By construction, $F_0$ is a continuous function on all of $\SLSL$.
This manifold has one cusp at $y\to\infty$.
The asymptotic behaviour is here
\eqn\smrt{
\eqalign{
F_0(M) \sim & \sum_{n\neq 0} f_0\bigg((0,ny^{-1/2})
\pmatrix{ \cos(\theta/2) & \sin(\theta/2) \cr
-\sin(\theta/2) & \cos(\theta/2) \cr} \bigg) \cr
\sim &
C_0(\theta) \, y^{1/2} \cr}
}
as $y \to \infty$,
where
\eqn\smrt{
\eqalign{
C_0(\theta) & = \int_{-\infty}^\infty f_0\bigg((0,r)
\pmatrix{ \cos(\theta/2) & \sin(\theta/2) \cr
-\sin(\theta/2) & \cos(\theta/2) \cr} \bigg) dr \cr
& =\frac{1}{\pi^2}\int_{-\infty}^\infty
\frac{e^{-\pi [r\sin(\theta/2)]^2} }{[r\cos(\theta/2)]^2} 
\chi_0(r\cos(\theta/2)) dr .\cr}
}
Note that $C_0(\theta)=O(1)$ for all $\theta$.

We re-write $F_1$ as a sum over primitive lattice points $\vecp$,
\eqn\proofi{
F_1(M)=\sum_{l=1}^\infty \sum_{\vecp} f_1(l\vecp M) .
}
For every primitive lattice point $\vecp$ there is a $K\in\SLZ$ such that
$\vecp=(0,1)K$. The subgroup $\Gamma_\infty\subset\SLZ$
of elements $K$ such that $(0,1)K=(0,1)$ is
\eqn\proofi{
\Gamma_\infty=\left\{ \pmatrix{ 1 & n \cr 0 & 1 \cr}:
n\in\ZZ \right\}
}
and hence there is a one-to-one correspondence between primitive lattice 
points and the coset $\Gamma_\infty\backslash\SLZ$.
We have therefore
\eqn\proofi{
F_1(M)=\sum_{l=1}^\infty \sum_{K\in\Gamma_\infty\backslash\SLZ} 
f_1((0,l) KM) .
}
Due to the rapid decay of $f_1$ this is essentially a finite sum.
To understand the singularities of $F_1$ consider the term corresponding
to $K=1$,
\eqn\proofi{
\sum_{l=1}^\infty f_1((0,l)M)
= \frac{y}{\cos(\theta/2)^2} 
\sum_{l=1}^\infty \frac{e^{-\pi l^2 y^{-1} \sin(\theta/2)^2}}{\pi^2 l^2} 
\chi_1(l y^{-1/2}\cos(\theta/2)).
}
The main singularity of this function is at $\theta=\pi$, and we note that
for $\theta \to \pi$
\eqn\proofi{
\sum_{l=1}^\infty f_1((0,l)M)
\sim \frac{4y}{\pi^2(\theta-\pi)^2} 
\sum_{l=1}^\infty \frac{e^{-\pi l^2 y^{-1}}}{l^2} .
}
The singularities of $F_1(M)$ are the images of the 
two-dimensional subspace $\{(z,\theta):\theta=\pi\}$
under the action of $\pmatrix{ a & b \cr c & d \cr}\in
\Gamma_\infty\backslash\SLZ$,
\eqn\sing{
\pmatrix{ a & b \cr c & d \cr}
\{(z,\theta):\theta=\pi\}
=
\{(z,\theta):\theta=\pi-2\arg(cz+d)\},
}
where $(c,d)$ runs over all primitive lattice points in $\ZZ^2-\{\vecnull\}$.

\subsec{Limit theorems}

Our main application of the above construction is the following.

\bigskip
{\bf Theorem 1}. There is a probability density $P(X)$ on $\RR_+$ with the following
properties:

\item{1.}
There is a set of $x$ of full measure such that,
for any bounded continuous function $\phi:\RR_+\to\RR$,
\eqn\proofi{
\lim_{T\to\infty}
\frac{1}{T} \int_0^T \phi(e^{-t/2}
g(e^{-t};x)) dt = \int_0^\infty \phi(X) P(X) dX;
}
\item{2.}
For any bounded continuous function $\phi:\RR_+\to\RR$, 
\eqn\proofi{
\lim_{y\to 0}
\int_0^1 \phi\bigl(\sqrt y g(y;x) \bigr) dx = \int_0^\infty \phi(X) P(X) dX ;
}
\item{3.}
As $X\to\infty$,
\eqn\proofi{
P(X) \sim A X^{-3/2} ,
}
with
\eqn\proofi{
A = \frac{3}{2\pi^3} \int_0^\infty  
\bigg(\sum_{l=1}^\infty \frac{e^{-\pi l^2 y^{-1}}}{l^2}\bigg)^{1/2} 
\frac{dy}{y^{3/2}} .
}

\bigskip

Note that the limiting distribution does not possess a first moment,
\eqn\proofi{
\int_0^\infty X P(X) dX =\infty.
}
Thus, there is no ``average'' value of $g_{cl}$ for Melvin models.

To prove the above limit theorem, we note that the
ergodicity of the geodesic flow and the equidistribution of long
closed horocycles on $\SLSL$ imply the following statements,
cf. \marklof. 

\bigskip
{\bf Theorem 2} 

\item{1.}
There is a set of $x$ of full measure such that,
for any bounded continuous function $G:\SLSL\to\CC$,
\eqn\proofi{
\lim_{T\to\infty}
\frac{1}{T} \int_0^T 
G\bigg(\pmatrix{ 1 & x \cr 0 & 1 \cr}\Phi^t\bigg) 
dt = \int_{\SLSL} G(M) dM;
}
\item{2.}
For any bounded continuous function $G:\SLSL\to\CC$, 
\eqn\proofi{
\lim_{y\to 0} \int_0^1
G\bigg(\pmatrix{ 1 & x \cr 0 & 1 \cr}
\pmatrix{ y^{1/2} & 0 \cr 0 & y^{-1/2} \cr}\bigg) 
dx = \int_{\SLSL} G(M) dM.
}
\bigskip

Now take any compactly supported continuous function $\phi:\RR_+\to\RR$
and set $G(M)=\phi(F(M))$. Then
for $\epsilon>0$ small enough
\eqn\proofi{
G(M)=\phi(F_0(M))
}
and hence $G(M)$ is 
bounded continuous, in view of the above singularity analysis.
Theorem 2 therefore implies the first two statements of Theorem 1, 
for compactly supported continuous test functions $\phi$, with
\eqn\proofi{
P(X)=\int_{\SLSL} \delta(X-F(M)) dM .
}
The extension to bounded continuous $\phi$ follows from a standard
probabilistic argument.

\subsec{Tail estimates}

Consider first the large $X$ asymptotics of
\eqn\smrt{
\eqalign{
P_1(X) & =\int_{\SLSL} \delta(X-F_1(M)) dM \cr
& =\int_{\SLSL} 
\delta\left(X-\sum_{K\in\Gamma_\infty\backslash\SLZ} 
\sum_{l=1}^\infty  f_1((0,l) KM)\right) dM \cr
& \sim \int_{\SLZ\backslash U_\sigma} 
\delta\left(X-\sum_{K\in\Gamma_\infty\backslash\SLZ} 
\sum_{l=1}^\infty  f_1((0,l) KM)\right) dM ,\cr}
}
where $U_\sigma=
\SLZ \{(z,\theta): \theta\in\pi+[-\sigma,\sigma]\}$ 
is a small neighbourhood 
of the singular set \sing\  on which $F_1(M)$ is large. 
Since the $K$ sum is essentially
finite, we may choose $\sigma>0$ small enough so that the overlap of
the neighbourhoods $K\{(z,\theta): \theta\in\pi+[-\sigma,\sigma]\}$ 
for different $K$ is negligible. Hence for $X\to \infty$,
\eqn\smrt{
\eqalign{
P_1(X) & \sim \sum_{K\in\Gamma_\infty\backslash\SLZ} 
\int_{\SLZ\backslash U_\sigma} 
\delta\left(X-
\sum_{l=1}^\infty  f_1((0,l) KM)\right) dM \cr
& = \int_{\Gamma_\infty\backslash U_\sigma} 
\delta\left(X-
\sum_{l=1}^\infty  f_1((0,l) M)\right) dM \cr
& = \frac{3}{2\pi^2} \int_0^1\int_0^\infty \int_{\pi-\sigma}^{\pi+\sigma} 
\delta\left(X-
\sum_{l=1}^\infty  f_1((0,l) M)\right) \frac{dx\,dy\,d\theta}{y^2} \cr
& \sim \frac{3}{2\pi^2} \int_0^\infty \int_{-\sigma}^{\sigma} 
\delta\left(X-
\frac{4 h(y)}{\theta^2} \right)
\frac{dy\,d\theta}{y^2} \cr
& = \frac{3}{2\pi^2 X^{3/2}} \int_0^\infty \sqrt{h(y)}
\frac{dy}{y^2} \cr}
}
where
\eqn\proofi{
h(y)= \frac{y}{\pi^2} \sum_{l=1}^\infty \frac{e^{-\pi l^2 y^{-1}}}{l^2}.
}
Since $F_0(M)$ has its only singularity in the cusp $y\to\infty$,
\eqn\smrt{
\eqalign{
P_0(X) & =\int_{\SLSL} \delta(X-F_0(M)) dM \cr
& \sim \frac{3}{2\pi^2}
\int_0^{2\pi} \int_0^\infty \delta(X-C_0(\theta)y^{1/2})
\frac{dy\,d\theta}{y^2} \cr
& = \frac{3}{\pi^2 X^3} \int_0^{2\pi} C_0(\theta)^2 d\theta .\cr}
}
So $P(X)\sim P_1(X)$ for large $X$, and the proof of Theorem
Theorem 1 is complete.

\listrefs

\bye